\documentclass[runningheads]{llncs}
\usepackage{graphicx}
\usepackage{amsmath}
\begin{document}

\title{Multi-task Swin Transformer for Motion Artifacts Classification and Cardiac Magnetic Resonance Image Segmentation}
\titlerunning{Multi-task Classification and Segmentation of CMR}
%
%
%
\author{Michal K. Grzeszczyk\inst{1} \and Szymon Płotka\inst{1, 2} \and Arkadiusz Sitek \inst{3}}
\authorrunning{M. K. Grzeszczyk et al.}

\institute{Sano Centre for Computational Medicine, Cracow, Poland \email{m.grzeszczyk@sanoscience.org} \and
Informatics Institute, University of Amsterdam, Amsterdam, The Netherlands \and
Massachusetts General Hospital, Harvard Medical School, Boston, MA, USA}
%
%
\maketitle              
\begin{abstract}
Cardiac Magnetic Resonance Imaging is commonly used for the assessment of the cardiac anatomy and function. The delineations of left and right ventricle blood pools and left ventricular myocardium are important for the diagnosis of cardiac diseases. Unfortunately, the movement of a patient during the CMR acquisition procedure may result in motion artifacts appearing in the final image. Such artifacts decrease the diagnostic quality of CMR images and force redoing of the procedure. In this paper, we present a Multi-task Swin UNEt TRansformer network for simultaneous solving of two tasks in the CMRxMotion challenge: CMR segmentation and motion artifacts classification. We utilize both segmentation and classification as a multi-task learning approach which allows us to determine the diagnostic quality of CMR and generate masks at the same time. CMR images are classified into three diagnostic quality classes, whereas, all samples with non-severe motion artifacts are being segmented. Ensemble of five networks trained using 5-Fold Cross-validation achieves segmentation performance of DICE coefficient of 0.871 and classification accuracy of 0.595.

\end{abstract}
\section{Introduction}
Cardiac Magnetic Resonance Imaging (CMR) is often used to assess cardiac anatomy and/or function \cite{lima2004cardiovascular}. As a noninvasive method, CMR gained popularity due to the generation of high-quality images enabling the diagnosis of multiple diseases of the human heart. Unfortunately, Magnetic Resonance Imaging (MRI) acquisition procedure is long and is susceptible to many artifacts \cite{smith2010mri}. Motion artifacts are prevalent in CMR due to the high blood flow in the field of view (FOV), breathing, or patient physical motion.

The delineation of the left ventricle (LV) and right ventricle (RV) blood pools together with left ventricular myocardium (MYO) can be useful for the diagnosis of cardiac diseases \cite{white1987left}. However, manual segmentation of cardiac structures is time-consuming and requires expert's knowledge. Thus, several methods for automated CMR segmentation have been proposed. In recent years Deep Learning (DL) allowed automated software to achieve human-level segmentation accuracy \cite{bernard2018deep}. Convolutional Neural Networks (CNNs) have been heavily applied for such tasks \cite{queiros2021right,al2021late}. Unfortunately, the performance of such models shown in many publications often cannot be matched when applied to data populations different from the one used for training.  The image properties differ between institutions due to different imaging procedures and equipment used for data acquisition. Motion artifacts are also important factors that prevent achieving  accurate segmentations. 

In this paper, we present a Multitask Swin UNEt TRansformers (Swin UNETR) network that enables the classification of motion artifacts  of the CMR acquired under different breathing conditions and simultaneous segmentation of LV, RV and MYO. We present the performance of the model on the CMRxMotion-challenge data organized alongside the Medical Image Computing and Computer Assisted Intervention (MICCAI) 2022 conference. 

The rest of this paper is structured as follows. In the next section, we present works related to multi-task learning and segmentation with DL. Section 3 describes our model and the dataset. In Section 4, results for tasks of motion artifacts classification and CMR segmentation are shown. 

\begin{figure}[t!]
    \centering
    \includegraphics[width=12cm]{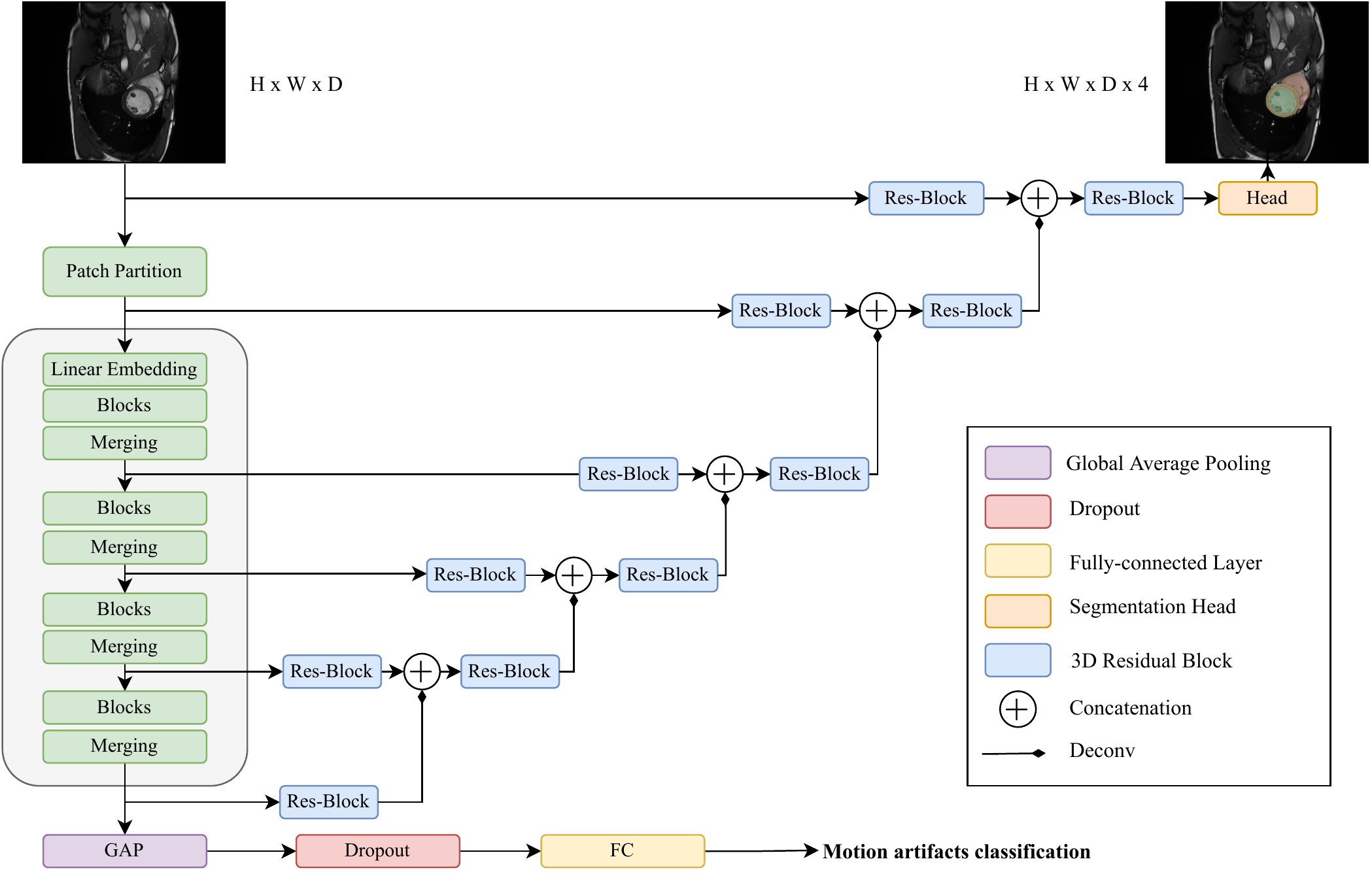}
    \caption{The overview of the multi-task Swin UNETR (inspired by \cite{hatamizadeh2022swin}) for CMR segmentation and diagnostic quality assessment. Swin Transformer is utilized as the backbone, encoder network of the U-Net architecture. The outputs of the encoder are passed to the convolutional decoder for segmentation purposes and to the classification branch for motion artifacts classification.}
    \label{fig:MSwinUnetr}
\end{figure}

\section{Related works}
Since the introduction of AlexNet in 2012 \cite{krizhevsky2012imagenet} deep neural networks (especially CNNs) have become the go-to tool for solving computer vision problems. One of the most popular CNN architecture utilized for image segmentation is U-Net \cite{ronneberger2015u}. U-Net was initially developed for biomedical image segmentation and comprises of CNN encoder and decoder with skip connections between different levels of those components. This approach was later improved and changed in various ways \cite{kossaifi2019t,ottom2022znet}.

Deep learning is also being applied in other domains. For example, Transformer architecture was proposed for the text translation task in the Natural Language Processing area \cite{vaswani2017attention}. Transformer-based models achieve state-of-the-art results on the language understanding benchmarks thanks to the utilization of the self-attention mechanism which allows for detecting long-range dependencies between elements of the sequence. This idea was later refined to the image domain when the Vision Transformer (ViT) was introduced \cite{dosovitskiy2020image}. ViT divides images into non-overlapping patches which are then fed into the Transformer Encoder. Unfortunately, ViT requires substantially big datasets to achieve high performance, has the quadratic computation complexity depending on the image size and lacks the inductive bias of CNNs. Therefore, Swin Transformer (Shifted Window Transformer) computing self-attention only within local windows which scales linearly was presented \cite{liu2021swin}. Additionally, Swin Transformer applies the \textit{shifted window} approach which changes the local window for self-attention computation in each layer.

To overcome the limitations of Transformer architectures while retaining their advantages, hybrid architectures combining CNNs with Transformers are being proposed \cite{hassani2021escaping,plotka2022babynet}. One of the examples is Swin UNETR model \cite{hatamizadeh2022swin}. In this architecture, a 3D Swin Transformer is used as the encoder of the U-Net-like architecture and multiple convolutional Residual Blocks (Res-Blocks) are utilized in the decoder part. In this paper we utilize Swin UNETR architecture as the base for our model.

Multi-task learning (MTL) is the area of DL whose target is to improve the performance of deep neural networks via simultaneous training of the network to solve multiple tasks \cite{zhang2018overview}. It has been shown that MTL can improve the model's performance even if the number of data samples is limited \cite{dobrescu2020doing}. In this paper, we present the utilization of MTL for motion artifacts classification and CMR segmentation.

\section{Method}

In this section, we present our approach to CMR segmentation on the CMRxMotion challenge dataset. We employ Multitask Swin UNETR to solve two tasks in this challenge: 1) motion artifacts classification, and 2) CMR segmentation.

\subsection{Multitask Swin UNETR}
To solve both problems simultaneously with one network, we employ a Multitask Swin UNETR (Fig. \ref{fig:MSwinUnetr}). We follow the architecture proposed by \cite{hatamizadeh2022swin}. Swin UNETR is similar to the U-Net architecture \cite{ronneberger2015u}. However, instead of utilizing fully convolutional DL model, Swin UNETR uses Swin Transformer as the encoder network. At first, the input to the Swin Transformer is partitioned into patches which are then projected into embedding space. Such tokens are passed to Swin Transformer blocks computing the following equations \cite{liu2021swin}: 

\begin{equation}
    \begin{split}
    \hat{z}^{l} &= WMSA(LN(z^{l-1})) + z^{l-1} \\ 
     & z^{l} = MLP(LN(\hat{z}^{l})) + \hat{z}^{l} \\
     & \hat{z}^{l+1} = SWMSA(LN(z^{l})) + z^{l} \\
     & z^{l+1} = MLP(LN(\hat{z}^{l+1})) + \hat{z}^{l+1}
    \end{split}
\end{equation}

Here, MLP is Multilayer Perceptron, LN denotes layer normalization and $z^{i}$ is the output from previous Swin Transformer blocks or the embedding layer in case of the first block. WMSA and SWMSA are ordinary and shifted-window multi-head self-attentions (MHSA), respectively. Swin Transformer blocks compute MHSA on \textit{$M \times M \times M$} windows (extracted from the input patches) instead of the full image, to enable linear computational complexity growth depending on the image size (ordinary self-attention computation on the full image has the quadratic computational cost). The SWMSA is computed with the shifted window approach where the windows for computing MHSA are shifted by ($\lfloor \frac{M}{2} \rfloor$, $\lfloor \frac{M}{2} \rfloor$, $\lfloor \frac{M}{2} \rfloor$) pixels. The self-attention is computed according to the following equation:
\begin{equation}
Attention(Q, K, V) = Softmax(\frac{QK^T}{\sqrt{d}})V,
\end{equation}
where \textit{Q, K, V} are queries, keys, values,  while \textit{d} denotes the \textit{Q} size. After, Swin Transformer blocks patches are merged again.

The Swin UNETR's decoder is created out of convolutional layers. To solve the motion artifacts assessment task, we use high-level feature map representation, and we fed them to the classification branch. The output of the decoder is global average pooled and passed through one Dropout and Fully-connected layers. We set the number of features at each Swin Transformer layer to \textit{k} $\times$ 60, where \textit{k} is the features factor at each layer.

\section{Experiments and results}

\subsection{Dataset}
CMRxMotion dataset's purpose is to investigate the impact of motion artifacts on the performance of automated CMR segmentation methods. It contains data from 45 volunteers (20 patients in train, 5 in validation and 20 in test datasets). Every volunteer underwent end-diastolic (ED) and end-systolic (ES)  acquisition of CMR under \textit{full breath-hold}, \textit{half breath-hold}, \textit{free breathing}, and \textit{intensive breathing} conditions. This results in 160, 40, and 160 images in the training, validation, and test sets. Different acquisition conditions created different levels of motion artifacts from the least in \textit{full breath-hold} to the most in \textit{intensive breathing}.  All images were labeled by experts into three classes: \textit{1} - mild motion, \textit{2} - intermediate motion, \textit{3} - severe motion. The \textit{1}, \textit{2} classes have sufficient quality for diagnosis. In all samples labelled as such LV, RV and MYO were segmented. The aim of the challenge is to create an algorithm for motion artifacts classification into three classes describing the severity of motion artifacts (task 1) and perform segmentation from samples from diagnostic quality samples \textit{1} and \textit{2} (task 2). Samples with severe motion artifacts were not used in the segmentation task. 

\subsection{Loss function}

The model is trained to minimize the sum of segmentation and classification losses defined as:
\begin{equation}
    \mathcal{L}_{sum} = \lambda_{1} (\mathcal{L}_{CE_{seg}} + \mathcal{L}_{Dice}) + \lambda_{2} \mathcal{L}_{CE_{cls}},
\end{equation}
where $\lambda_{1} = 2.25$ and $\lambda_{2} = 1$, $\mathcal{L}_{CE_{seg}}$ is Cross-Entropy loss calculated on the segmentation mask, $\mathcal{L}_{CE_{cls}}$ is loss calculated on the classification labels and $\mathcal{L}_{Dice}$ is DICE coefficient loss computed on non-background classes. The segmentation component of loss function is omitted for samples with severe motion artifacts.

\subsection{Implementation details}

All training samples are of shape $H \times W \times D$, where $H\in[400, 512]$ $W\in[594, 731]$ $D\in[12, 18]$. We resize all volumes to the same patch size of $256 \times 256 \times 32$ (the \textit{D} channel is padded with zeros). The output of the model is of shape $H \times W \times D \times 4$ (background class and three segmentation classes - LV, RV, MYO), where each slice in the final dimension is the segmentation class mask. We train five models using 5-Fold Stratified Cross-validation where the stratification groups are selected based on classification labels  to ensure that motion artifacts are evenly distributed across all folds. During training, we apply multiple spatial transformations to avoid overfitting. Among those augmentations there are: random flips for each axis (p = 0.2), random zoom (p = 0.1), random rotate (p = 0.1). Before the augmentations, we normalize the input data. The model is implemented with PyTorch and trained on $2 \times$ NVIDIA A100 80GB GPUs. We train the model for 250 epochs during each fold with AdamW optimizer. We set an initial learning rate to 2e-04 with the Cosine Annealing learning rate scheduler. We use the default parameters of Swin UNETR from the MONAI library \cite{monai2020monai} unless stated otherwise. 

\subsection{Evaluation metrics}

To measure the performance of the model we utilize the DICE coefficient and Hausdorff Distance 95 (HD95) as segmentation metrics as well as Accuracy and Cohen's Kappa as classification metrics.

\begin{figure}[t!]
    \centering
    \includegraphics[width=12cm]{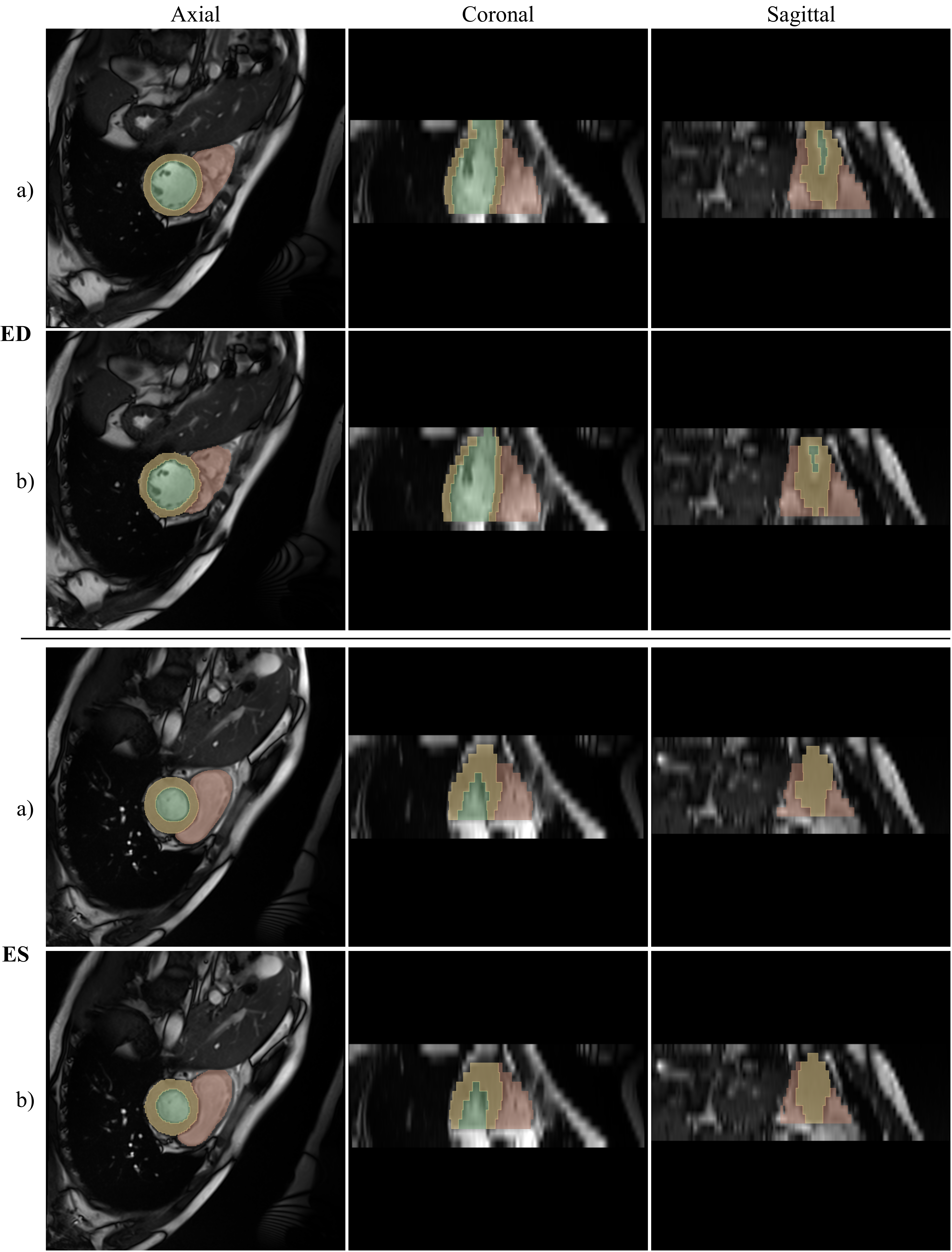}
    \caption{Results of the segmentation of end-diastolic (ED) and end-systolic (ES) images shown on the three CMR views (Axial, Coronal, Sagittal) a) ground truth, b) Swin UNETR.}
    \label{fig:segmentation}
\end{figure}

\subsection{Motion artifacts classification}
Motion artifacts classification is a difficult task due to the highly imbalanced dataset - 70 samples with mild motion, 69 with intermediate motion and only 21 with severe motion. The results achieved by Swin UNETR on different folds and validation set are presented in Table \ref{tab:class}. The model reaches up to 0.75 accuracy and 0.59 Cohen's Kappa on the third fold's test set. On average, the accuracy on all folds is 0.595 and Cohen's Kappa reaches 0.184. 
\bgroup
\def\arraystretch{1.5}%
\begin{table}[t!]
    \caption{Average Accuracy and Cohen's Kappa among folds on CMRxMotion challenge training and validation datasets}
    \begin{center}
        \begin{tabular}{c||c|c|c|c|c||c}
            \textbf{Metric} & \textbf{Fold 1} & \textbf{Fold 2} & \textbf{Fold 3} & \textbf{Fold 4} & \textbf{Fold 5} & \textbf{Average}\\
            \hline
            \textbf{Accuracy - KFold test} & 0.594 & 0.6 & 0.75 & 0.5 & 0.531 & 0.595\\
            \textbf{Accuracy - validation} & - & - & - & - & - & 0.525\\
            \textbf{Cohen's Kappa - KFold test} & 0.257 & 0.286 & 0.59 & 0.0 & 0.0 & 0.184\\
            \textbf{Cohen's Kappa - validation} & - & - & - & - & - & 0.127\\
        \end{tabular}
    \end{center}
    \label{tab:class}
\end{table}
\egroup

\bgroup
\def\arraystretch{1.5}%
\begin{table}[t!]
    \caption{Average Dice and Hausdorff distance among folds on CMRxMotion challenge training and validation datasets}
    \begin{center}
        \begin{tabular}{c||c|c|c|c|c||c}
            \textbf{Metric} & \textbf{Fold 1} & \textbf{Fold 2} & \textbf{Fold 3} & \textbf{Fold 4} & \textbf{Fold 5} & \textbf{Average}\\
            \hline
            \textbf{Dice - KFold test} & 0.88 & 0.858 & 0.898 & 0.89 & 0.831 & 0.871\\
            \textbf{Dice - validation} & - & - & - & - & - & 0.807\\
            \textbf{HD95 - KFold test} & 2.245 & 6.347 & 1.794 & 3.271 & 12.187 & 5.169\\
            \textbf{HD95 - validation} & - & - & - & - & - & 9.803\\
        \end{tabular}
    \end{center}
    \label{tab:seg}
\end{table}
\egroup

\subsection{CMR segmentation}
The exemplary results of CMR segmentation are presented in Figure \ref{fig:segmentation}. Swin UNETR manages to delineate RV, LV and MYO with high accuracy. The biggest decrease in the DICE coefficient arises from the lack of a smooth mask and issues with small details in the output. Segmentation metrics are presented in Table \ref{tab:seg}. Swin UNETR achieves an average DICE coefficient of 0.871 on all folds with HD95 of 5.169.

\section{Conclusions}
In this paper we presented a Multi-task Swin UNETR network for CMRxMotion challenge tasks solving. The proposed model was evaluated on the CMR dataset. This network is able to perform motion artifacts classification and CMR segmentation in a single forward pass. Such an approach enables the utilization of more data samples in the single network training (which enhances network's capabilities) and the application of only one model for solving two tasks which is important due to lower computational complexity (contrary to the utilization of two networks). In the future, the model's performance can be improved by utilizing more data (for example from other CMR segmentation challenges like ACDC \cite{bernard2018deep}).

\section*{Acknowledgements}

This work is supported by the European Union’s Horizon 2020 research and innovation programme under grant agreement Sano No 857533 and the International Research Agendas programme of the Foundation for Polish Science, co-financed by the European Union under the European Regional Development Fund.

%
%
%
\bibliographystyle{splncs04}
\bibliography{bibliography}

\end{document}